\documentclass[conference]{IEEEtran}
\IEEEoverridecommandlockouts
\usepackage{cite}
\usepackage{amsmath,amssymb,amsfonts}

\usepackage{caption} 
\usepackage{tabularx}
\usepackage{float}
\usepackage{array}  
\usepackage{booktabs} 
\usepackage{multirow}
\usepackage{makecell}

\usepackage{color}

\usepackage{bbding}

\usepackage{amsfonts,amssymb} 

\usepackage{subfigure} 

\usepackage{algorithmic}
\usepackage{graphicx}
\usepackage{textcomp}
\usepackage{xcolor}
\def\BibTeX{{\rm B\kern-.05em{\sc i\kern-.025em b}\kern-.08em
    T\kern-.1667em\lower.7ex\hbox{E}\kern-.125emX}}
\begin{document}


\title{Relation-guided acoustic scene classification aided with event embeddings}


\author{


\IEEEauthorblockN{Yuanbo Hou}
\IEEEauthorblockA{
\textit{WAVES} \\
\textit{Ghent University}\\
Gent, Belgium \\
Yuanbo.Hou@UGent.be}
\and
\IEEEauthorblockN{Bo Kang}
\IEEEauthorblockA{
\textit{IDLAB} \\
\textit{Ghent University}\\
Gent, Belgium \\
Bo.Kang@UGent.be}
\and
\IEEEauthorblockN{Wout Van Hauwermeiren}
\IEEEauthorblockA{
\textit{WAVES} \\
\textit{Ghent University}\\
Gent, Belgium \\
Wout.VanHauwermeiren@UGent.be}
\and
\IEEEauthorblockN{Dick Botteldooren}
\IEEEauthorblockA{
\textit{WAVES} \\
\textit{Ghent University}\\
Gent, Belgium \\
Dick.Botteldooren@UGent.be
}

}

\maketitle

\begin{abstract}
In real life, acoustic scenes and audio events are naturally correlated. Humans instinctively rely on fine-grained audio events as well as the overall sound characteristics to distinguish diverse acoustic scenes. Yet, most previous approaches treat acoustic scene classification (ASC) and audio event classification (AEC) as two independent tasks. A few studies on scene and event joint classification either use synthetic audio datasets that hardly match the real world, or simply use the multi-task framework to perform two tasks at the same time. Neither of these two ways makes full use of the implicit and inherent relation between fine-grained events and coarse-grained scenes. To this end, this paper proposes a relation-guided ASC (RGASC) model to further exploit and coordinate the scene-event relation for the mutual benefit of scene and event recognition. The TUT Urban Acoustic Scenes 2018 dataset (TUT2018) is annotated with pseudo labels of events by a simple and efficient audio-related pre-trained model PANN, which is one of the state-of-the-art AEC models. Then, a prior scene-event relation matrix is defined as the average probability of the presence of each event type in each scene class. Finally, the two-tower RGASC model is jointly trained on the real-life dataset TUT2018 for both scene and event classification. The following results are achieved. 1) RGASC effectively coordinates the true information of coarse-grained scenes and the pseudo information of fine-grained events. 2) The event embeddings learned from pseudo labels under the guidance of prior scene-event relations help reduce the confusion between similar acoustic scenes. 3) Compared with other (non-ensemble) methods, RGASC improves the scene classification accuracy on the real-life dataset.

\end{abstract}


\section{Introduction}
\label{sec:intro}
Acoustic scene classification (ASC) tags audio recordings using the predefined semantic labels that characterize the environment and situation in which it was recorded.  
Audio event classification (AEC) is dedicated to multi-label classification on audio clips and aims to identify the presence of target audio events.
ASC and AEC can be used in a wide variety of applications such as robot hearing \cite{ren2016sound}, audio forensics \cite{malik2013acoustic}, emergency detection \cite{principi2015integrated} and road surveillance \cite{almaadeed2018automatic}.

Prior studies related to IEEE AASP Challenges in Detection and Classification of Acoustic Scenes and Events (DCASE) \cite{DCASE2016, DCASE2017, DCASE2018} commonly handle ASC and AEC as two separate tasks and tune models for each task individually. However, real-world audio streams include both acoustic
scenes and events and they are inherently  correlated. 
For example, in the acoustic scene \textit{metro station}, audio events of \textit{bell ringing} and \textit{engine starting} are likely to occur. Such fine-grained events are the fundamental building blocks of polyphonic acoustic scenes. 
Therefore, a joint scene and event recognition method based on an artificially synthesized dataset  is proposed in \cite{Bear2019TowardsJS}, expecting to train a shared acoustic feature encoder for scenes and events.
However, the artificial dataset in \cite{Bear2019TowardsJS} does not accurately catch complexities between real-world acoustic scenes and events.
Then in \cite{chandrakala2019generative}, robust representations for environmental audio scenes and events are learned by generative model-driven representations and have proved to be effective in audio-related tasks. 
Another class of studies for joint analysis of scene and event refers to multi-task learning (MTL) \cite{jung2021dcasenet}.  
Several convolutional layers are shared in a multi-task model as they \cite{tonami2021joint} expect to learn and utilize shared low-level representations and separated high-level representations of scenes and events. 
However, the one-hot hard labels of scenes used in \cite{tonami2021joint} cannot model the extent to which audio events and acoustic scenes are related. 
To alleviate this issue, the output of a trained scene model is used as the teacher in \cite{imoto2020sound} to guide the learning of the scene branch, which works as the student, in the joint scene-event classification model based on the teacher-student learning \cite{t-s}.
To learn the knowledge of events under scene conditions, 
a scene-event joint analysis model based on scene-conditioned loss is proposed \cite{komatsu2020scene}.
Overall, in previous scene and event joint analysis works, papers \cite{Bear2019TowardsJS}\cite{chandrakala2019generative}\cite{tonami2021joint} do not explore the implicit scene-event relation,
paper \cite{imoto2020sound} exploits the one-way scene-to-scene relation.
Although paper \cite{komatsu2020scene} uses the one-way scene-to-event relation by conditional loss, that relation is derived from the artificial dataset in \cite{Bear2019TowardsJS}, and is difficult to match the complex and intricate scene-event relation in the real world.


In contrast to prior works, this paper is not only interested in obtaining a shared representation encoder with scenes and events knowledge, but also in how the real-world implicit and inherent scene-event relation can be used to guide the model to bidirectionally fuse the information of coarse-grained scenes and fine-grained events to reduce confusion of similar scenes, even if the event information is derived from unverified pseudo labels (a proxy to unavailable ground-truth labels) \cite{pseudolabel}.

To the best of our knowledge, there are no publicly available large real-life datasets that contain both acoustic scene and event labels. Hence, a large real-life acoustic scene dataset,  TUT Urban Acoustic Scenes
2018\protect\footnote{Dataset available:  https://zenodo.org/record/1228142\#.YfJ-Qv7MJnI}, with diverse audio events is used in this paper \cite{DCASE2018}. In order to obtain labels for events in the real-life acoustic scene dataset, a simple and efficient pre-trained audio-related model PANN \cite{kong2020panns} is used to tag each audio clip with pseudo labels of 527 classes of audio events. 
Relying on true labels of scenes and pseudo labels of events, a scene-event relation matrix is derived to model the implicit relation between real-life scenes and events.
Then, with the aid of pseudo labels and the prior knowledge of joint scene-event relation matrix, a relation-guided ASC and AEC two-tower model is proposed to mutually estimate the knowledge of scenes and corresponding events and explore the possibility of collaboratively classifying scenes and events. 

This paper is organized as follows. Section 2 introduces the method. Section 3 describes the dataset, experimental setup, results, and analysis. Finally, Section 4 gives conclusions.

\section{Method}
In this section, the scene-event relation matrix is presented and is applied to the relation-guided two-tower convolutional neural  networks (CNN) for the acoustic scene classification (ASC) and audio event classification (AEC) tasks.

\subsection{Prior scene-event relation matrix }

Coarse-grained acoustic scenes are highly correlated with fine-grained audio events, for example, \textit{car passing} and \textit{horn screeching} are more likely to occur in the \textit{street} scene than \textit{cheerful music}. 
In contrast, \textit{joyful music} accompanied by  \textit{people walking} is more common in the \textit{shopping mall} scene. 
The relation between real-life scenes and events is not simply 1 for presence or 0 for absence, but rather a likelihood expressed by probability.
Inspired by such an intuitive observation, this paper attempts to build the scene-event relation matrix on the real-world acoustic scene dataset, instead of the simple connections in \cite{komatsu2020scene} represented by 0 and 1 on synthetic datasets.  
Since there are no public large real-life datasets that contain both acoustic scenes and events labels, 
a lightweight and effective audio-related model PANN \cite{kong2020panns} is used in this paper to tag audio clips with pseudo labels of events. PANN \cite{kong2020panns} is trained and performs well on Audioset \cite{aduioset}, which contains 527 classes of polyphonic audio events in daily life.

Given the probability of 527 classes of events for the $i$-th audio clip ($a_i$) of the $j$-th scene $S_j$ is $P(S_j, a_i) \in \mathbb{R}^{1\times 527}$,
\begin{equation}
\setlength{\abovedisplayskip}{3pt}
\setlength{\belowdisplayskip}{3pt}
P(S_j, a_i) = [p_{e_1}, p_{e_2}, p_{e_3}, ..., p_{e_{527}}]
\end{equation}
where $p_{en}\in [0,1], n \in [1,527]$, and 
$p_{e_n}$ is the probability of the occurrence of the $n$-th event in the audio clip.
The $p_{en}\in [0,1]$ implies that in the AEC task, different audio events are generally considered to be independent of each other.
Then, the average probability of audio events in the acoustic scene $S_j$ can be notated as $P(S_j)$,

\begin{equation}
\setlength{\abovedisplayskip}{3pt}
\setlength{\belowdisplayskip}{3pt}
P(S_j) = 1 / I_j \sum\nolimits_{i=1}^{I_j} P(S_j, a_i)
\end{equation}
where $I_j$ is the total number of audio clips in the scene $S_j$.
Then, the prior scene-event relation matrix ${R}_{SE}$ is composed of $P(S_j)$ rows for all scenes: ${R}_{SE} \in \mathbb{R}^{K\times 527}$, 
\begin{equation}
\setlength{\abovedisplayskip}{3pt}
\setlength{\belowdisplayskip}{3pt}
R_{SE} = [P(S_1), P(S_2), P(S_3), ..., P(S_K)]^T
\end{equation}
where $K$ is the number of acoustic scenes in the dataset.
Next, $R_{SE}$ from the training set will be introduced into the two-tower model of ASC and AEC. 
The role of $R_{SE}$ is to guide the model to coordinate and utilize the implicit relation between coarse-grained scenes and fine-grained events during the training phase. 

\subsection{Relation-guided collaborative two-tower model}\label{sec:architecture}
\begin{figure}[t]
	\centerline{\includegraphics[width = 0.5\textwidth]{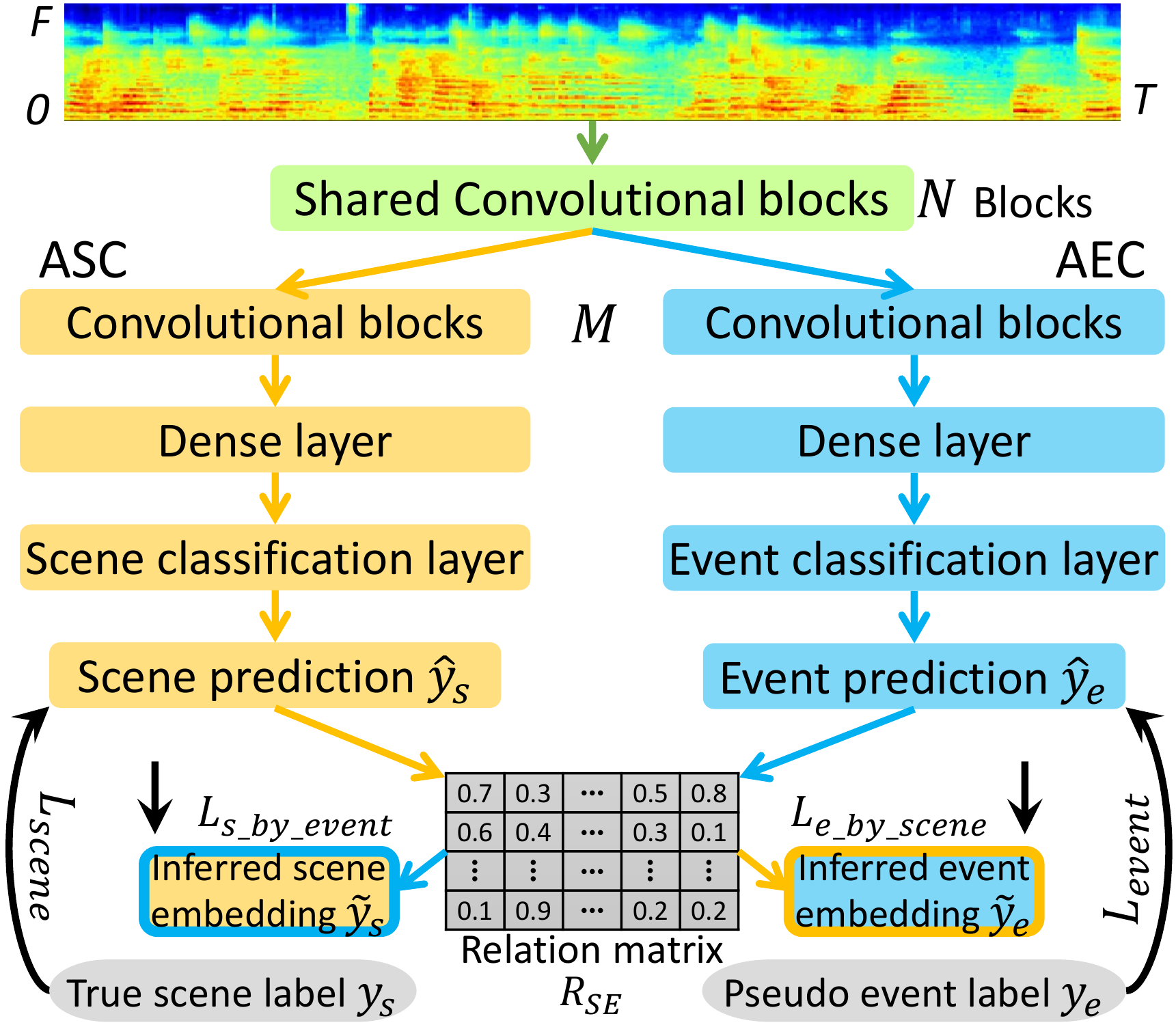}}
	\caption{The architecture of the relation-guided acoustic scene classification (RGASC) model.}
	\label{model}
\end{figure}

The core of this part is how to embed the existing fixed prior relation matrix $R_{SE}$ relating coarse-grained information from true labels of scenes and fine-grained information from pseudo labels of events into the learning process of the model. To exploit the fixed prior knowledge of $R_{SE}$, a relation-guided two-tower model is proposed in Fig. \ref{model}.

\label{ssec:model}

The raw waveform is first converted to time-frequency (T-F) representations using log mel spectrograms \cite{mel}.
Then, the first $N$ convolutional blocks of the two-tower model are used to extract shared basic acoustic features of scene and event as inspired by \cite{Bear2019TowardsJS}, \cite{chandrakala2019generative} and \cite{tonami2021joint}.
Compared with high-level acoustic features, basic local acoustic features learned by models are more transferable and applicable \cite{interspeech2020hyb}, which is beneficial for model generalization \cite{generalization}. Then, the remaining $M$ convolutional blocks are applied to the ASC tower and the AEC tower to capture local patterns that are beneficial to scenes and events, respectively.
There are a total of 6 convolutional blocks in the proposed model. 
Therefore, given the first $N$ convolution blocks are shared, the remaining $M=6-N$ convolutional blocks will be used to learn the task-oriented representations of each tower.
The optimal ratio of $N$ and $M$ will be further explored in the experimental section.

Referring to CNNs in VGG \cite{vgg}, each convolutional block contains 2 convolutional layers with the kernel size of (3 × 3). 
Batch normalization \cite{batchnormal} and ReLU activation functions \cite{relu} are used to accelerate and stabilize the training.
Next, a linear dense layer is applied to the high-level representations, followed by a scene classification dense layer with softmax activation function and an event classification dense layer with sigmoid activation function, respectively.
 More details and code,
please see the {homepage}\protect\footnote{{Homepage: https://github.com/Yuanbo2020/RGASC}}.

ASC is a task dedicated to single-label multi-class classification, so the cross entropy loss \cite{DCASE2018} is used as the loss function in the ASC tower between the scene prediction $\hat{y}_{s}\in\mathbb{R}^{K}$ and the scene true  label ${y}_{s}\in\mathbb{R}^{K}$,
\begin{equation}
\setlength{\abovedisplayskip}{3pt}
\setlength{\belowdisplayskip}{3pt}
L_{scene} = - \sum\nolimits_{j=1}^{K} 
{y}_{s_j}\log(\hat{y}_{s_j})
\end{equation}
AEC aims to perform multi-label classification on audio clips to detect multiple targets simultaneously. Therefore, the binary cross-entropy \cite{kong2020panns} loss is used in the AEC tower between the event prediction $\hat{y}_{e}\in\mathbb{R}^{527}$ and the pseudo label ${y}_{e}\in\mathbb{R}^{527}$,
\begin{equation}
\setlength{\abovedisplayskip}{3pt}
\setlength{\belowdisplayskip}{3pt}
    L_{event} = - \sum\nolimits_{i=1}^{527} 
{y}_{e_i}\log(\hat{y}_{e_i}) + (1-{y}_{e_i})\log(1-\hat{y}_{e_i})
\end{equation}

To guide the model to explore relations between scenes and events based on the prior knowledge of $R_{SE}$ from the training set.
The scene prediction $\hat{y}_{s}$ is mapped to the latent event space via $R_{SE}$ to obtain the corresponding event information $\Tilde{y}_{e}$ inferred from the scene prediction, 
$\Tilde{y}_{e}=\hat{y}_{s}\cdot R_{SE}$.
The $R_{SE}$ is derived from the probability of events in each scene, hence the embedding vector $\Tilde{y}_{e}$ uses the prior information of events in the scene.
To measure the distance between the inferred event vector $\Tilde{y}_{e}$ and the actual event prediction $\hat{y}_{e}$ in the latent space,
the $\hat{y}_{e}$ from the AEC tower is used as the reference in the mean squared error (MSE) loss,
\begin{equation}
\setlength{\abovedisplayskip}{3pt}
\setlength{\belowdisplayskip}{3pt}
     L_{\text{e\_by\_scene}} = 1/ 527 \sum\nolimits_{i=1}^{527} 
(\hat{y}_{e_i}-\Tilde{y}_{e_i})^2 
\end{equation}
The embedding vector $\Tilde{y}_{e}$ is not a probability distribution, so the regression loss MSE is used to minimize the relation-guided $L_{\text{e\_by\_scene}}$ loss of inferred event by scene information. This expects the inferred event vector $\Tilde{y}_{e}$ to be close to the actual event vector $\hat{y}_{e}$ in the latent representation space.
Furthermore, the study \cite{mse_ce} on the entropy-based loss and MSE shows that MSE loss can better correct the error between the estimated value and the target. 

Similarly, the inner product $\Tilde{y}_{s}=\hat{y}_{e}\cdot R_{SE}^T$ defines the relation-guided embedding vector for scenes.
The relation-guided embedding vector $\Tilde{y}_{s}$ from event prediction indicates the possibility of different scenes, and higher similarity means the corresponding scene is more likely to occur.
Similar to $\Tilde{y}_{e}$, $\Tilde{y}_{s}$ is not a probability distribution, 
so MSE is adopted to minimize the loss of inferred scene by event information 
to expect the inferred scene vector $\Tilde{y}_{s}$ to be close to the actual scene vector $\hat{y}_{s}$ in the latent representation space.
\begin{equation}
\setlength{\abovedisplayskip}{3pt}
\setlength{\belowdisplayskip}{3pt}
     L_{\text{s\_by\_event}} =1 / K \sum\nolimits_{i=1}^{K} 
(\hat{y}_{s_i}-\Tilde{y}_{s_i})^2
\end{equation}

The final loss function of the two-tower models is given by the weighted sum of the separate loss functions:
\begin{equation}
\setlength{\abovedisplayskip}{3pt}
\setlength{\belowdisplayskip}{3pt}
L = \lambda_1L_{\text{scene}} + \lambda_2L_{\text{s\_by\_event}} + \lambda_3L_{\text{event}} + \lambda_4L_{\text{e\_by\_scene}}
\end{equation}
where $\lambda_i$ is the scale factor of each loss function. $\lambda_i$ defaults to 1.
In the experimental section, various configurations of $\lambda_i$ are explored. 
The ASC tower will benefit from $L_{\text{scene}}$ and $L_{\text{e\_by\_scene}}$.
The loss $L_{\text{e\_by\_scene}}$
is fed to the ASC tower, expecting to obtain a more coordinated scene prediction with the event prediction of AEC tower.
Likewise, the AEC tower will benefit from $L_{\text{event}}$ and $L_{\text{s\_by\_event}}$, 
the loss $L_{\text{s\_by\_event}}$ benefits the collaborative learning of AEC tower.

\section{Experiments and results}

\subsection{Dataset and Experimental Setup}
The dataset used in this paper is the TUT Urban Acoustic Scenes 2018 development dataset   (TUT2018) \cite{DCASE2018} with 8640 10-seconds clips totaling 24 hours and contains 10 classes of acoustic scenes from real life. For each acoustic scene, there are 864 examples. These audio recordings were recorded in 6 different European locations. This real-life dataset does not contain labels for events. Therefore, to obtain the event labels, a lightweight pure CNN-based pre-trained model PANN\protect\footnote{{We used the CNN14 in PANN in this paper. The pre-trained model (the model named Cnn14\_16k\_mAP=0.438.pth) of CNN14 is available here: https://zenodo.org/record/3987831}}  \cite{kong2020panns} is used to tag each audio clip with a pseudo label indicating the probability of 527 classes of audio events. 

The log mel-bank energy with 64 banks \cite{mel} is used as the acoustic feature in this paper. This is extracted by the Short-Time Fourier Transform (STFT) with a Hamming window length of 46 \textit{ms} and a window overlap of $1/3$ \cite{Kong2018}. 
Dropout \cite{dropout} and normalization are used in the training to prevent over-fitting of the model. 
Adam optimizer \cite{adam} with a default initial learning rate of 0.001 minimizes the loss function.
The default training batch size is 64. 
To facilitate the comparison of experimental results with other systems, the training/testing split of the TUT2018 dataset follows the default split of the DCASE 2018 Task 1 Subtask A\protect\footnote{http://dcase.community/challenge2018/task-acoustic-scene-classification}.
The model is trained on a single graphical card (Tesla V100-SXM2-32GB) for a fixed amount of 100 epochs.
The average accuracy (Acc) \cite{Kong2018} is used as the metric in this paper. A higher Acc indicates a better performance to distinguish different scenes.

\subsection{Results and Analysis}
This section analyzes the performance of the proposed method based on the following \textbf{R}esearch \textbf{Q}uestions (\textbf{RQ}):

\noindent
\textbf{• RQ1}: How many convolutional blocks should be shared?

\label{ssec:layers}
\begin{figure}[t]
	\centerline{\includegraphics[width = 0.5\textwidth]{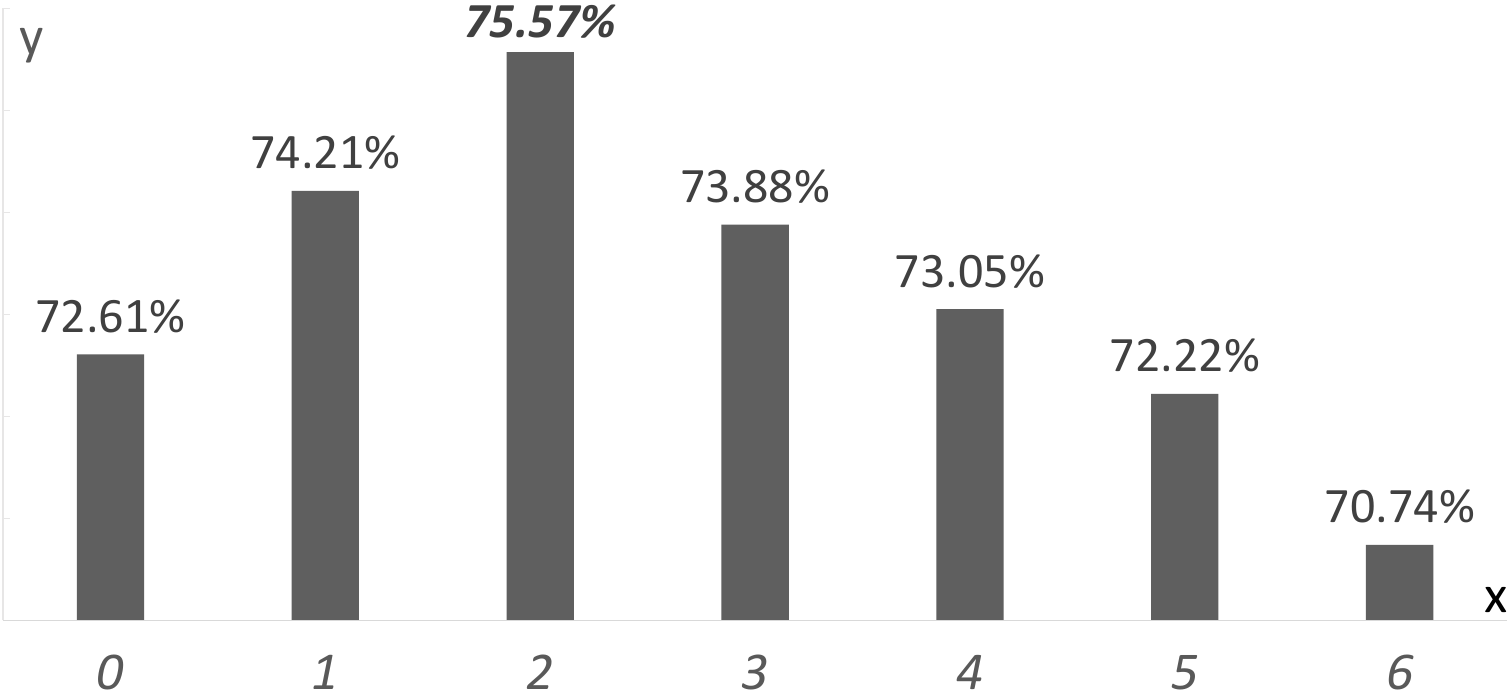}}
	\caption{The effect of different numbers of shared blocks on the proposed model on the test set of real-life dataset. X-axis is the number of shared blocks, and y-axis is the accuracy of scene classification.}
	\label{layers}
\end{figure}

As discussed in Section \ref{sec:architecture}, the first question to be explored is what proportion of the joint scene-event representations should be shared? 
From another perspective, the ratio between shared and individual blocks determines whether the model is dedicated to learning shared knowledge or is more inclined to explore task-dependent knowledge. There is both competition and mutual influence between the shared and separated blocks in the RGSAC model.

There are a total of 6 convolution blocks in the proposed RGASC model. 
Therefore, when the number of shared blocks is 0 in Fig. \ref{layers}, the learning of scene representations and event representations in the two-tower RGASC model will be independent, and the model will not be able to learn the joint scene-event representation.
When the number of shared blocks is 6, the learning of scene and event representations will be completely overlapping, then the model can only rely on the subsequent dense layer of each tower to learn the individual task-oriented representations.

As shown in Fig. \ref{layers}, increasing the number of shared blocks does not consistently improve the classification accuracy of the model. The performance of the model peaks when the number of shared blocks is 2. 
Then, further increase in the number of shared blocks will 
degrade 
the performance of the model.
That means the optimal structure of the two-tower model is the structure when the number of shared blocks is 2. Subsequent experiments will be conducted on this model structure.

\begin{table}[!b]   
	\renewcommand\tabcolsep{1pt} 
	\centering
	\caption{The ablation study of the proposed model.}
	\begin{tabular}
	{p{0.8cm}<{\centering}
	p{1.5cm}<{\centering}
	p{1.5cm}<{\centering}
	p{1.5cm}<{\centering}
	p{1.5cm}<{\centering}
	p{1.4cm}<{\centering}}
	
		\toprule[1pt] 
		\specialrule{0em}{0.1pt}{0.1pt}
	  {\#} & $L_{\text{scene}}$ & $L_{\text{s\_by\_event}}$ & $L_{\text{event}}$ & $L_{\text{e\_by\_scene}}$ & \textsl{Acc (\%)}\\
		\midrule[1pt]  
		\specialrule{0em}{0.2em}{0.pt}
		1 & \CheckmarkBold & \XSolidBrush  & \XSolidBrush  &  \XSolidBrush &  71.46 \\ 
		\specialrule{0em}{0.pt}{0.pt}
		2 & \XSolidBrush  & \XSolidBrush & \CheckmarkBold  &  \XSolidBrush &  36.90\\
		\specialrule{0em}{0.pt}{0pt}
		3 & \CheckmarkBold &  \XSolidBrush & \CheckmarkBold  & \XSolidBrush & 72.94 \\ 
		
		\specialrule{0em}{0pt}{0em}
	4 & \XSolidBrush& \CheckmarkBold &  \XSolidBrush & \CheckmarkBold  & 10.07 \\
		\specialrule{0em}{0pt}{0em}
		
			\specialrule{0em}{0pt}{0em}
	5 & \CheckmarkBold &  \XSolidBrush & \CheckmarkBold  & \CheckmarkBold&  \textbf{75.80}  \\
		\specialrule{0em}{0pt}{0em}
		
			\specialrule{0em}{0pt}{0em}
	6 & \CheckmarkBold &  \CheckmarkBold & \CheckmarkBold  & \XSolidBrush& 74.36 \\
		\specialrule{0em}{0pt}{0em}
		\bottomrule[1pt]
	\end{tabular}
	\label{tab:ablation}
\end{table}

\noindent
\textbf{• RQ2}: How do different weights $\lambda_i$ for the four losses influence the performance of the model?

During the training phase, different values of $\lambda_i$ represent the difference in  importance of scene information ($L_{\text{scene}}$), scene information inferred from the event prediction ($L_{\text{s\_by\_scene}}$), event information ($L_{\text{event}}$), and event information inferred from the scene prediction ($L_{\text{e\_b\_event}}$), respectively. Except for the scene information, 
the other three types of information are derived from pseudo labels.
Since only the labels of scenes are available in the real-life dataset TUT2018, where the accuracy of the pseudo labels of events tagged by the pre-trained model PANN \cite{kong2020panns} cannot be evaluated due to the absence of reference event labels, and the goal of this paper is to improve the accuracy of scene classification, so the experiments will focus on the results of ASC.

In this section, the importance of different types of information is explored that corresponds to different $\lambda$ for model learning. First, an ablation study has been conducted to compare the role of different types of information in the proposed RGASC. 
Table \ref{tab:ablation} lists the result of enabling and disabling certain parts of RGASC.
For \# 1 in Table \ref{tab:ablation}, only scene information is exploited, which is a pure ASC (puASC) model without the aid of additional information. 
\# 2 predicts scenes by $\Tilde{y}_{s}=\hat{y}_{e}\cdot R_{SE}^T$, which calculates the similarity between each row of the relation matrix $R_{SE}^T$ and event information $\hat{y}_{e}$ provided by pseudo labels to derive the possibility of each scene.
The $R_{SE}^T$ with fixed prior knowledge can be viewed as a table.
That is, \# 2 in Table \ref{tab:ablation} can be regarded as the result of a lookup table that relies on
the event information from pseudo labels, so its result is poor. \# 3 uses the true information of scenes and pseudo information of events to obtain a shared joint scene-event representations extractor, which is similar to prior works \cite{Bear2019TowardsJS}\cite{chandrakala2019generative}\cite{tonami2021joint}.
\# 4 relies only on the prior relation matrix to derive each other's outputs without learning the knowledge of scenes from the true labels and the knowledge of events from the pseudo labels. 
Without the support of accurate task-dependent representations of scenes and events, \# 4, which only learns the implicit and intricate scene-event relation through the prior relation matrix, actually cannot learn the knowledge of scene classification, resulting in its inability to classify effectively, so its performance is poor.
Compared with \# 3, \# 5 based on $L_{\text{e\_by\_scene}}$ expects to obtain more accurate and coordinated event-related scene prediction and enhance the discrimination ability of the scene branch,  which in turn brings a better classification result.
In contrast to \# 5, \# 6 attempts to obtain more accurate scene-related event information.
\# 5 outperforms \# 6, which indicates that increasing the weight of scene information is more beneficial to the ASC task.
 The ablation study in Table \ref{tab:ablation} shows that the more the model pays attention to the scene-related information, the better its performance.

\begin{table}[b]\footnotesize
	\renewcommand\tabcolsep{1.5pt} 
	\centering
\caption{The effect of different $\lambda_i$ values on the ASC task.}
	\begin{tabular}
	{
	p{0.85cm}<{\centering}|
	p{1.4cm}<{\centering}
	p{1.4cm}<{\centering}
	p{1.4cm}<{\centering}
	p{1.4cm}<{\centering}|
	p{1.5cm}<{\centering}
	} 
		\hline
		{\#} & 
		$\lambda_1$ &  $\lambda_2$ & $\lambda_3$ &  $\lambda_4$ & 
		\textsl{Acc (\%)}  \\
\hline
     1 & 1 &  0 &  1 &  0.01  & 75.92  \\
     2  &   1 &   0 &   0.5 &   0.01 & 76.27   \\
     3 &  1 &   0.1 &   1 &   0.1 & 75.05\\
      4 &  1 &   0.1 &   0.1 &   0.1 & 75.62 \\
    5  &   1 &   0.1 &   0.5 &   0.1  & 75.79  \\
     6  &   1 &   0.01 &   0.5 &   0.01  & \textbf{\textsl{77.35}} \\
      7  &    1 &   0.01 &   0.01 &   0.01 & 76.48\\
      8  &    1 &   0.001 &   0.01  &   0.001 & 76.71 \\
		
\hline

	\end{tabular}
	\label{tab:loss}
\end{table}

Second, fine-grained control of the weight of each loss function is explored.
The fusion of different semantic information in Table \ref{tab:loss} tries to adjust the weights of the other three kinds of information from pseudo labels to maximize their benefit. 
In other words, the right amount of noise-filled pseudo information needs to be introduced to help recognize similar scenes. 
Finally,
giving maximum weight to $L_{\text{scene}}$ and secondary weight to $L_{\text{event}}$, 
while absorbing the scene-event relation information
($L_{\text{s\_by\_event}}$ and $L_{\text{e\_by\_scene}}$) with weaker weights, makes the best result of \# 6 in Table \ref{tab:loss}, which gradually merges the coarse-grained
true information of scene and the fine-grained pseudo information of event based on scene-event relation matrix.

\begin{table}[t]  
	\renewcommand\tabcolsep{1.5pt} 
	\centering
	\caption{Comparison of classification results of different systems on the development set of TUT2018.}
	\begin{tabular}{
	p{3.2cm}<{\centering}|
	p{3.9cm}<{\centering}|
	p{1.0cm}<{\centering}
	}
	
	    \hline
	    
		System & Model structure & \textsl{Acc (\%)}\\
		
		\hline
		PANN \cite{kong2020panns} (Fixed mode) & VGG-like CNN & 56.9 \\
			
		Baseline \cite{DCASE2018} & CNN & 59.7 \\

		CNN\_Surrey \cite{Kong2018} & CNN &  68.0 \\
		
		NNF\_CNNEns \cite{Nguyen2018a} & CNN and nearest neighbor filters & 69.3 \\

		Attention \cite{wang2018self} & CRNN with Self-attention & 70.8 \\
		
	 	ABCNN \cite{Ren2018} & Attention-Based CNN & 72.6 \\
		
		PANN (Fine-tuning mode) & VGG-like CNN & 73.8 \\
		
		MLTF \cite{Zhang2018-svm} & CNN and SVM & 75.3 \\

		Wavelet-based DSS \cite{Li2018} & CRNN & 76.6 \\

		Proposed RGASC & VGG-like CNN & \textbf{77.4} \\

	\hline
	\end{tabular}
	\label{tab:rule}
\end{table}

\noindent
\textbf{• RQ3}:
Does the proposed RGASC system in this paper perform better than other systems?

The published results of the DCASE2018 Task 1 Subtask A challenge are compared in this section in Table \ref{tab:rule}. 
Only the non-ensemble methods are taken into consideration\footnote{The non-ensemble results are obtained for comparison from the DCASE2018 Task 1 Subtask A (T1A) website. The proposed RGASC only uses a single model with one type of acoustic feature and does not involve any data augmentation methods, while the top 3 methods in T1A are mostly an ensemble of multiple models with multiple features, so the top 3 results are omitted in Table \ref{tab:rule}. In detail, the Top-1 in T1A is an ensemble of 2 deep CNN trained with 11 types of acoustic features. Next, the Top-2 uses depth-wise separable CNN trained with 3 multi-scale acoustic features. Finally, the Top-3 is an ensemble of 6 big and deep models trained with 4 types of acoustic features. Therefore, the ensemble was performed on up to 24 models in total.}.
In addition to the well-performing convolutional neural networks (CNN) \cite{DCASE2018}\cite{Kong2018}, 
convolutional recurrent neural network (CRNN) \cite{wang2018self} with self-attention works well, where self-attention \cite{attention} is used to model the relationship between different positions of sequences output by CRNN.
 Next, to achieve more optimal classification, an attention pooling layer is used in CNN to reduce the feature dimension \cite{Ren2018}.
 On the other hand, in addition to the exploration of feature dimensions, the paper \cite{Zhang2018-svm} proposes a CNN-based multiple layer temporal feature (MLTF) to try to capture the dynamic temporal information of audio signals efficiently.
 Furthermore, wavelet-based Deep Scattering Spectra (DSS) \cite{Li2018} is used to exploit higher-order information of acoustic features in the scene classification based on a CRNN model. 
For a comprehensive comparison, the lightweight and efficient pre-trained model PANN \cite{kong2020panns} used in this paper is also added to the comparison, and the performance of PANN is explored in two modes referring to the transfer learning \cite{interspeech2020hyb}. 
In fixed mode, the parameters of PANN will not be updated during training, it will use the prior knowledge of 527 classes of events learned from Audioset to classify scenes. In fine-tuning mode, PANN will learn and absorb scene information based on the existing knowledge to update parameters.

Results in Table \ref{tab:rule} show that the event knowledge has a certain ability to distinguish scenes. 
The score of fixed-mode PANN is close to Baseline, while the fine-tuned PANN gets a better result than the CRNN with self-attention. 
This indicates that even simple pure CNN without recurrent layers and diverse attention mechanisms can achieve promising results with the help of a large dataset (e.g. Audioset totals 5.8 thousand hours). 
Compared to other submissions in Table \ref{tab:rule}, 
the proposed RGASC for recognizing coarse-grained acoustic scenes aided by relation-guided fine-grained event information is effective,
even if the fine-grained event information is derived from pseudo labels without any verification.

\begin{table}[t]\normalsize
	\renewcommand\tabcolsep{1.5pt} 
	\centering
	\caption{Comparison of ASC results with other scene-event joint analysis methods on the development set of TUT2018.}
	\begin{tabular}{
	p{0.5cm}<{\centering}|
	p{6.8cm}<{\centering}|
	p{1.1cm}<{\centering}
	}
	    \hline
		\# & Method &  \textsl{Acc(\%)}\\
		\hline
		1 & Joint scene and event recognition \cite{Bear2019TowardsJS} &   52.35 \\ 
		
		2 & Joint event and scene analysis using MTL  \cite{tonami2021joint} &    61.69 \\
		
		3 & Conditional scene and event recognition \cite{komatsu2020scene} &   66.39 \\ 
		
		4 & RGASC &  \textbf{77.35} \\
	
	\hline
	\end{tabular}
	\label{tab:asc_joint}
\end{table}

Table \ref{tab:asc_joint} also compares the RGASC with some existing methods for joint scene-event analysis. 
Among them, the first listed method \cite{Bear2019TowardsJS} performs scene and event classification based on the same joint embedding space and scores the worst. 
This is easy to understand because real-life coarse-grained scenes and fine-grained events contain their own different characteristics and attributes. 
Then, the second-worst model \cite{tonami2021joint} based on MTL \cite{jung2021dcasenet} attempts to exploit both shared joint and separate individual representations of scenes and events.
The third method \cite{komatsu2020scene} jointly analyses scenes and events based on the one-way scene-to-event conditional loss. 
The performance of \cite{komatsu2020scene} is better than that of \cite{tonami2021joint}, which indicates that the scene-conditioned loss plays the expected role.
Overall, the proposed RGASC scores best out of the discussed joint analysis models of scenes and events.

\noindent
\textbf{• RQ4}:
Does the introduction of event information from pseudo labels improve the recognition of acoustic scenes?

	 

\begin{figure*}[t]
	\centering  
	\subfigure[Representations from puASC.]{\includegraphics[width=2.05in]{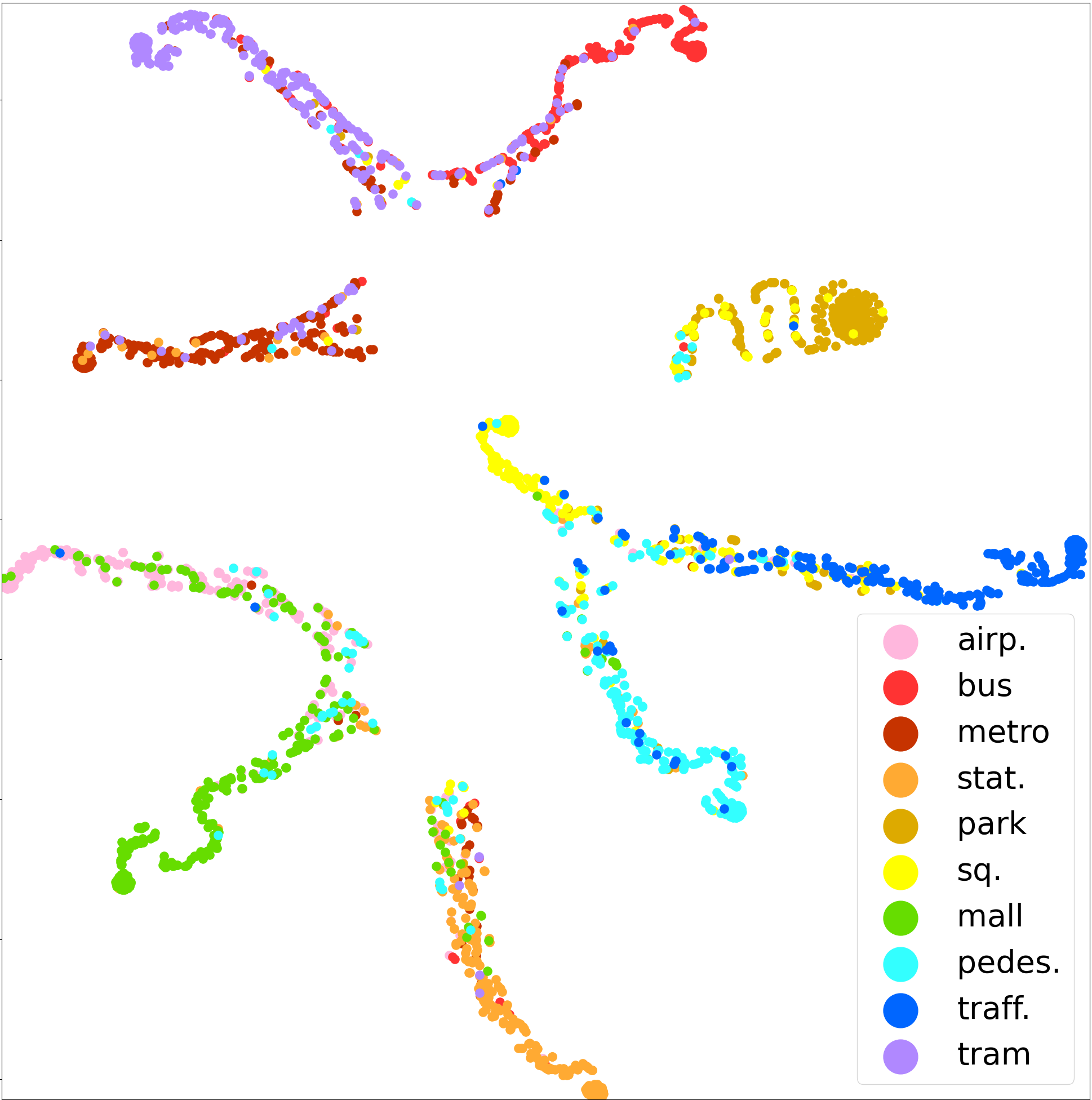} }\label{fig_first_case}
	\subfigure[Representations from RGASC.]{\includegraphics[width=2.05in]{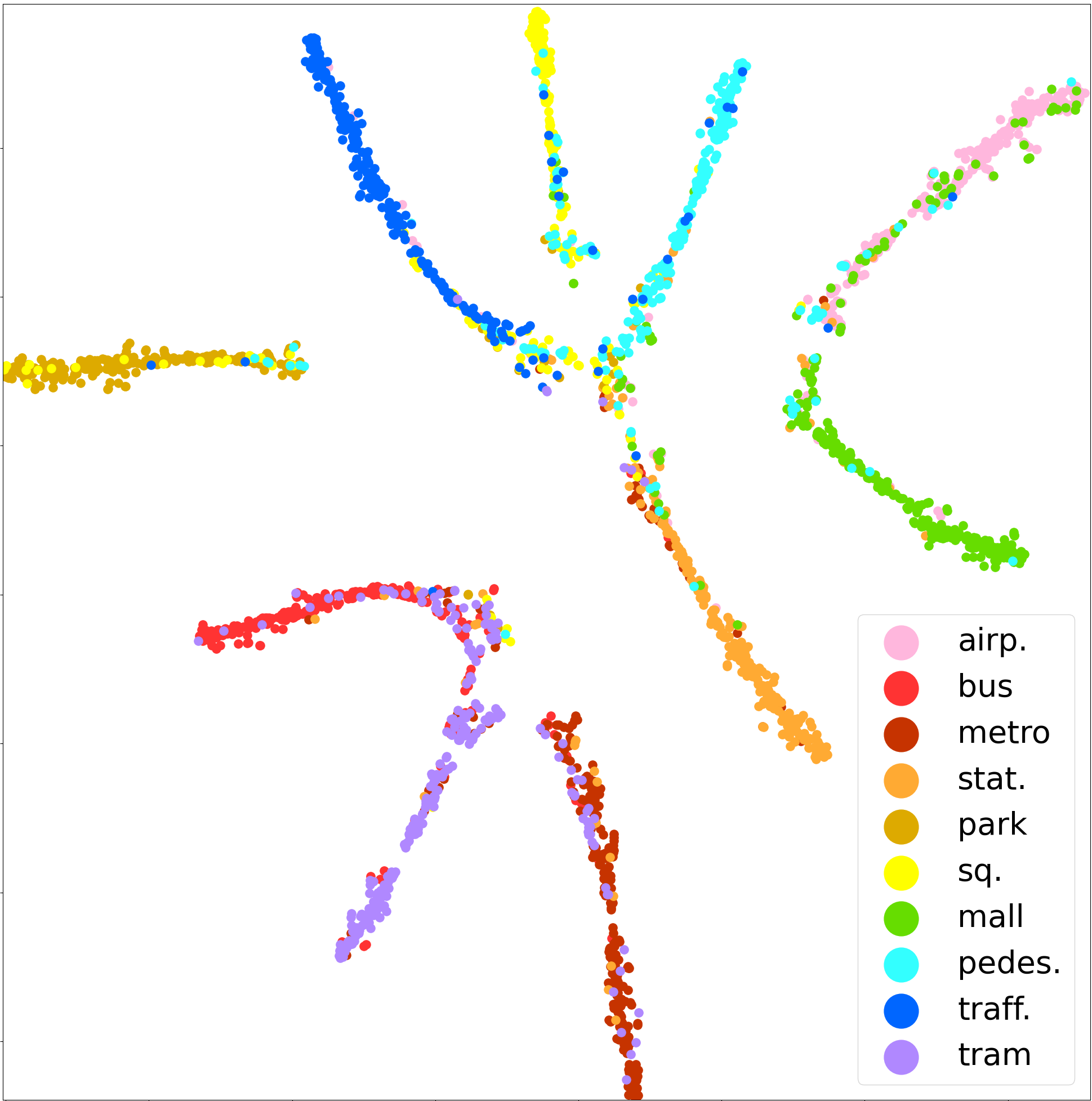}}
	 
	\subfigure[Confusion matrix of puASC.]{\includegraphics[width=2.09in]{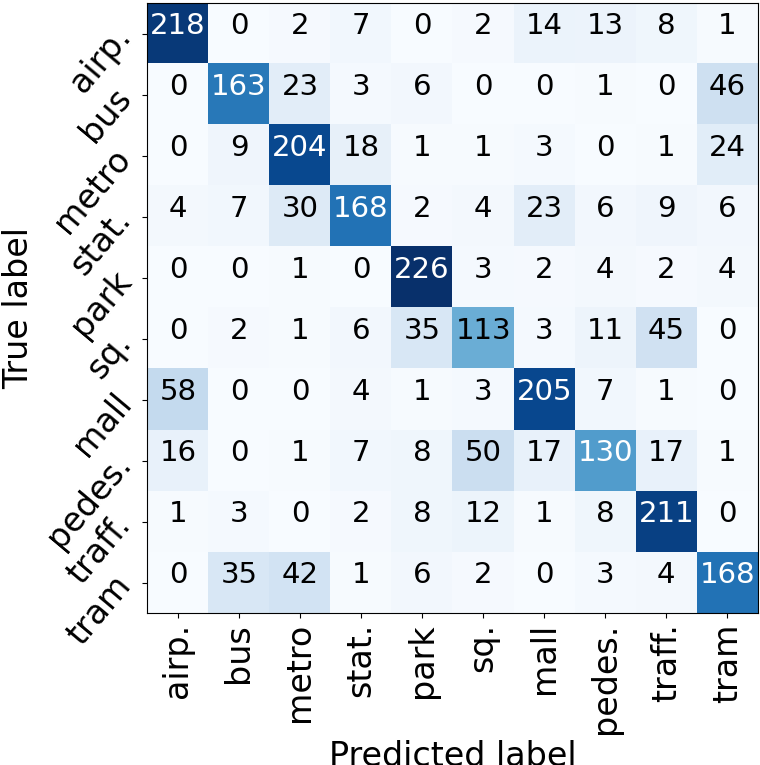} }\label{fig_first_case}\subfigure[Confusion matrix of RGASC.]{\includegraphics[width=2.09in]{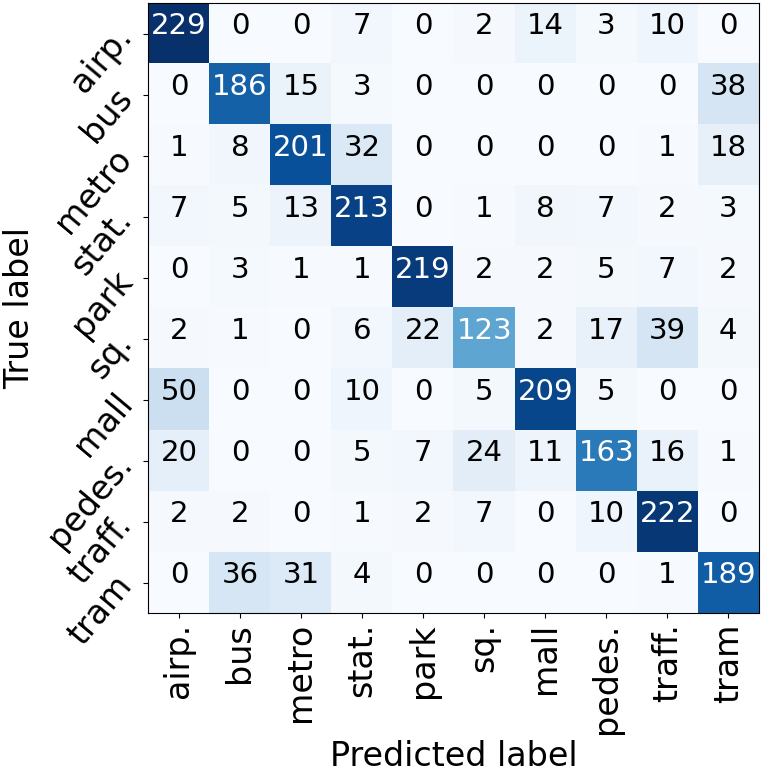}}
	\caption{Visualization of high-level representations of acoustic scenes from models of puASC and RGASC using t-SNE \cite{tsne} and the corresponding confusion matrices on the test set of the development set of TUT2018.}
	\label{tsne}
\end{figure*}


Table \ref{tab:scene} specifically shows the role of pseudo-label information.
  The accuracy of every scene class is compared for puASC (\# 1 in Table \ref{tab:ablation}) and the proposed RGASC (\# 6 in Table \ref{tab:loss}). 
  RGSAC effectively improves the classification accuracy except for the two scenes of \textit{park} and travelling by an underground metron (\textit{metro}). 
  In particular, the classification accuracy of the metro station (\textit{stat.}) and pedestrian street (\textit{pedes.}) is improved by 17.38\% and 13.36\%, respectively.

\begin{table}[b]\footnotesize
	\renewcommand\tabcolsep{2.0pt} 
	\centering
	\caption{Classification accuracy (\%) on test set of the puASC with only true labels of scenes and
	the proposed RGASC aided by pseudo labels of events.}
	\begin{tabular}
	{
	p{0.9cm}<{\centering} |
	p{0.61cm}<{\centering} |
	p{0.61cm}<{\centering} |
	p{0.61cm}<{\centering} |
	p{0.61cm}<{\centering} |
	p{0.61cm}<{\centering} |
	p{0.61cm}<{\centering} |
	p{0.61cm}<{\centering} |
	p{0.7cm}<{\centering} |
	p{0.61cm}<{\centering} |
	p{0.61cm}<{\centering}
	} 
		   
		\toprule[1pt]
		\specialrule{0em}{0pt}{0pt}
		
		\textit{scene} & 
		\textit{airp.} & 
		\textit{bus} & 
		\textit{metro} & 
		\textit{stat.} & 
		\textit{park} & 
		\textit{sq.} & 
		\textit{mall} & 
		\textit{pedes.} & 
		\textit{traff.} &
		\textit{tram} \\
		
		\hline
		
		\textit{puASC} & 
		\textit{82.26} &
		\textit{67.35} & 
		\textit{78.16} & 
		\textit{64.86} & 
		\textit{93.38} & 
		\textit{52.31} & 
		\textit{73.47} & 
		\textit{52.63} & 
		\textit{85.77} & 
		\textit{64.36} \\
		

		
		\hline
		
		\textit{RGASC} & 
		\textit{86.42} & 
		\textit{76.86} & 
		\textit{77.01} & 
		\textit{\textbf{82.24}} & 
		\textit{90.50} & 
		\textit{56.94} & 
		\textit{74.91} & 
		\textit{\textbf{65.99}} & 
		\textit{90.24} & 
		\textit{72.41} \\ 

		\specialrule{0em}{0pt}{0pt}
		\bottomrule[1pt]
		
	\end{tabular}
	\label{tab:scene}
\end{table}

To gain deeper insights, Fig. \ref{tsne} intuitively visualizes the gain of relation-guided pseudo-label information using t-SNE \cite{tsne}.
There are 9 sub-clusters in Fig. \ref{tsne} (a) in the 10-class classification task. 
Many samples from similar scenes, like \textit{bus} and \textit{tram}, public square (\textit{sq.}) and \textit{park}, street traffic (\textit{traff.}) and \textit{park}, are mixed.
Even for the human auditory system, relying on audio alone to distinguish these similar scenes is challenging \cite{complex_acoustic_scenes}.
The 10 classes of scenes are clearly shown in Fig. \ref{tsne} (b) of RGASC. Even the \textit{sq.} that are covered by other scenes in Fig. \ref{tsne} (a) are delineated as a separate sub-cluster.
The distinction between different scenes represented by different color sub-clusters is more obvious and the confusion is thus reduced, achieving a better classification result.
This indicates that the relation-guided information fusion between the fine-grained event pseudo labels and the coarse-grained scene true labels works. In addition, in the bottom row of Fig. \ref{tsne} the corresponding confusion matrices are presented.

The pseudo-labels of events used in this paper cannot ensure their accuracy due to the lack of ground-truth reference labels.
However, the experimental results show that the imprecise pseudo-label information introduced by the relation matrix does boost the accuracy of scene classification.
Pseudo labels without manual verification may prove their  benefits, because pseudo labels can still depict the possibility of events in different scenes to some extent \cite{pseudo-representation}.
Therefore, the AEC tower in Fig. \ref{model} trained based on pseudo labels can learn inaccurate but still effective event representations and distribution information, and transform this information into cues that can be applied to identify different scenes under the guidance of the relation matrix proposed in this paper, so as to enhance the recognition ability of the scene classification model.
From the perspective of teacher-student learning \cite{teacher-transfer}, the pre-trained model PANN \cite{kong2020panns} used to output event pseudo labels in this paper can be regarded as the teacher model, and the AEC tower aiming to extract event information based on pseudo labels can be regarded as the student model. 
The teacher model outputs its rich knowledge of events into pseudo labels and transfers it to the student model \cite{teacher2}. Although the student model may not perform as well as the teacher model, the student model still has some discernment about events \cite{transfer}.
In this paper, the fine-grained event information learned by the student model will be used as the reference to correct the event information inferred by the ASC tower to enhance the scene branch's discrimination of diverse events within the scene, and further identify the differences within different scenes to boost the discrimination of the ASC tower for similar events. And then, the accuracy of scene classification is improved.

\section{CONCLUSION}
\label{sec:CONCLUSION}

Inspired by natural relations between real-life varied acoustic scenes and diverse events, 
this paper proposes to use the scene-event relation to guide the model to collaboratively classify scenes and events.
The proposed relation-guided ASC (RGASC) framework
effectively
coordinates
the coarse-grained true information of scenes and fine-grained pseudo information of events.
Experiments show that the introduction of pseudo-label information performs well under the guidance of fixed prior relation matrix $R_{SE}$,
and RGASC shows promising performance in differentiating similar polyphonic scenes.

Future work will enable the model to autonomously learn the $R_{SE}$ during the training phase of model, and test it on more diverse datasets.

\section{ACKNOWLEDGEMENTS}
\label{sec:ACKNOWLEDGEMENTS}
This research received funding from the Flemish Government under the “Onderzoeksprogramma Artificiële Intelligentie (AI) Vlaanderen” programme.


\label{sec:refs}

\bibliographystyle{IEEEbib}
\bibliography{conference_101719}

\end{document}